\documentclass[12pt]{article}
\pdfoutput=1

\usepackage{a4wide}
\usepackage{amsmath,amssymb}
\usepackage{slashed}
\usepackage{xcolor}
\usepackage{graphicx}
\usepackage{url}
\usepackage{cite}
\usepackage[colorlinks=true,allcolors=darkred,pdfborder={0 0 0},linktocpage=false]{hyperref}
\usepackage{booktabs}

\definecolor{darkred}{rgb}{0.6,0,0}
\def\hc{\text{h.c.}}
\def\BR{\text{BR}}

\begin{document}

\vspace*{-2cm}
\begin{flushright}
IFIC/19-34 \\
\vspace*{2mm}
\end{flushright}

\begin{center}
\vspace*{15mm}

\vspace{1cm}
{\Large \bf 
Higgs lepton flavor violating decays in Two Higgs Doublet Models
} \\
\vspace{1cm}

{\bf Avelino Vicente}

 \vspace*{.5cm} 
Instituto de F\'{\i}sica Corpuscular (CSIC-Universitat de Val\`{e}ncia), \\
Apdo. 22085, E-46071 Valencia, Spain

\end{center}

\vspace*{10mm}
\begin{abstract}\noindent\normalsize
The discovery of a non-zero rate for a lepton flavor violating decay mode of the Higgs boson would definitely be an indication of New Physics. We review the prospects for such signal in Two Higgs Doublet Models, in particular for Higgs boson decays into $\tau \mu$ final states. We will show that this scenario contains all the necessary ingredients to provide large flavor violating rates and still be compatible with the stringent limits from direct searches and low-energy flavor experiments.
\end{abstract}

\newpage

\tableofcontents

\newpage

\section{Introduction} \label{sec:intro}

The discovery of the Higgs boson in 2012 at the LHC by the ATLAS and
CMS collaborations constitutes a historical milestone for particle
physics and another brilliant triumph for the Standard Model
(SM). With this long-awaited completion of the SM particle spectrum,
it stands as one of the most successful theories ever built, providing
precise predictions for a wide range of particle physics phenomena, in
good agreement with a large amount of experimental results at low and
high energies.

Despite this remarkable success, many fundamental questions remain
unanswered in the SM. The list is long and contains experimental
observations that cannot be addressed in the SM and theoretical issues
that cannot be fully understood in its context. It includes the origin
of neutrino masses, the nature of dark matter, the conservation of CP
in the strong interactions or the reason for the replication of
fermion generations, to mention a few. These open problems clearly
call for an extension of the SM with new states, presumably present at
high energies, and/or new dynamics.

New Physics (NP) may manifest in the form of Higgs boson properties
different from those predicted by the SM. For this reason, it is
crucial to look for deviations in the Higgs couplings to fermions and
gauge bosons and the Higgs boson total decay width or the existence of
new Higgs decay channels. In particular, new degrees of freedom
coupling to the SM leptons and the Higgs boson could induce non-zero
lepton flavor violating (LFV) Higgs boson decays, such as $h \to
\ell_i \ell_j$ with $i \ne j$, indeed a common feature in many models
with extended scalar or lepton sectors. The observation of these
processes, strictly forbidden in the SM, would provide a clear hint of
NP at work.

Higgs lepton flavor violating (HLFV) signatures face many indirect
constraints, since most of the NP scenarios that lead to them modify
other Higgs properties as well, some already experimentally determined
to lie close to the SM prediction.  Moreover, HLFV signatures
typically come along with other LFV processes, such as the $\ell_i \to
\ell_j \gamma$ radiative decays. While model-independent studies
\cite{Goudelis:2011un,Blankenburg:2012ex,Harnik:2012pb,Dery:2013rta,Celis:2013xja,Dery:2014kxa}
have shown that large HLFV rates are in principle compatible with the
existing experimental constraints, this is not generally the case in
specific models. In fact, most models predict HLFV rates below the
current LHC sensitivity. In contrast, the Two Higgs Doublet Model
(2HDM) \cite{Lee:1973iz,Branco:1980sz} has been shown to be able to
accommodate large $h \to \ell_i \ell_j$ branching ratios
\cite{Sierra:2014nqa}, clearly within the reach of the ATLAS and CMS
detectors.

This \textit{minireview} focuses on HLFV in the 2HDM. Several pioneer
works already addressed HLFV in the pre-LHC era
\cite{Pilaftsis:1992st,DiazCruz:1999xe,Sher:2000uq,DiazCruz:2002er,Brignole:2003iv,DiazCruz:2004tr,Kanemura:2004cn,Arganda:2004bz,Parry:2005fp},
and many have revisited the subject in the context of the 2HDM
\cite{Davidson:2010xv,Dery:2013aba,Kopp:2014rva} or in other contexts
\cite{DiazCruz:2008ry,Giang:2012vs,Arhrib:2012mg,Arhrib:2012ax,Arana-Catania:2013xma,Falkowski:2013jya,Arganda:2014dta}
after the LHC has started delivering data. In fact, early results by
the CMS collaboration hinted at a non-zero $h \to \tau \mu$ branching
ratio~\cite{Khachatryan:2015kon}, and this raised a considerable
attention in the community, leading to many
works~\cite{Sierra:2014nqa,Heeck:2014qea,Crivellin:2015mga,deLima:2015pqa,Dorsner:2015mja,Omura:2015nja,Vicente:2015cka,Das:2015zwa,Yue:2015dia,Bhattacherjee:2015sia,Kobayashi:2015gwa,Mao:2015hwa,He:2015rqa,Crivellin:2015hha,Altmannshofer:2015esa,Cheung:2015yga,Arganda:2015naa,Botella:2015hoa,Liu:2015oaa,Huang:2015vpt,Baek:2015mea,Baek:2015fma,Arganda:2015uca,Aloni:2015wvn,Benbrik:2015evd,Zhang:2015csm,Hue:2015fbb,Bizot:2015qqo,Buschmann:2016uzg,Han:2016bvl,Chang:2016ave,Belusca-Maito:2016axk,Chen:2016lsr,Alvarado:2016par,Banerjee:2016foh,Hayreter:2016kyv,Huitu:2016pwk,Chakraborty:2016gff,Lami:2016mjf,Thuc:2016qva,Baek:2016kud,Altmannshofer:2016oaq,Baek:2016pef,Herrero-Garcia:2016uab,Phan:2016ouz,Hammad:2016bng,Demidov:2016gmr,Wang:2016rvz,DiIura:2016wbx,Tobe:2016qhz,Aoki:2016wyl,Guo:2016ixx,Choudhury:2016ulr,Arganda:2017vdb,Herrero-Garcia:2017xdu,Thao:2017qtn,Chamorro-Solano:2017toq,Chakraborty:2017tyb,Qin:2017aju,Nguyen:2018rlb,Babu:2018uik,Hou:2019grj,Nomura:2019dhw}. We
will first follow some general model-independent arguments that
identify the 2HDM as a scenario with potentially large HLFV signatures
and later highlight some selected phenomenological results on HLFV in
the 2HDM. Even though the theoretical discussion will be general and
not concentrate on any particular combination of lepton flavors, we
will focus on $\tau \mu$ LFV in the phenomenological discussion.

The rest of the manuscript is structured as follows. In
Sec.~\ref{sec:exp} the current experimental bounds for the rates of
several LFV processes, including $h \to \ell_i \ell_j$, are briefly
discussed. Sec.~\ref{sec:eft} contains general model-independent
considerations, while Sec.~\ref{sec:2HDM} argues that a type-III 2HDM
may accommodate sizable $h \to \ell_i \ell_j$ branching ratios and
makes this point explicitly after introducing our 2HDM notation and
conventions. Finally, we comment on some 2HDM scenarios that generate
neutrino masses and may lead to large HLFV rates in
Sec.~\ref{sec:numass} and conclude in Sec.~\ref{sec:conclusions}.

\section{Experimental status}
\label{sec:exp}

\begin{table}[tb]
  \begin{center}
    \begin{tabular}{c c c}
      \toprule
      HLFV Decay BR & ATLAS & CMS \\ 
      \midrule
      $h \to \mu e$ & $6.1 \times 10^{-5}$~\cite{ATLAS:2019xlq} & $3.5 \times 10^{-4}$~\cite{Khachatryan:2016rke}\\ 
      $h \to \tau \mu$ & $2.8 \times 10^{-3}$ ~\cite{Aad:2019ugc} & $2.5 \times 10^{-3}$~\cite{Sirunyan:2017xzt}\\ 
      $h \to \tau e$ & $4.7 \times 10^{-3}$~\cite{Aad:2019ugc} & $6.1 \times 10^{-3}$~\cite{Sirunyan:2017xzt} \\ 
      \midrule
    \end{tabular}
    \begin{minipage}{0.9\linewidth}
      \caption{Experimental $95~\%$ C.L. upper bounds on HLFV
        branching ratios from the ATLAS and CMS collaborations.
        \label{tab:HLFVsignals}}
    \end{minipage}
  \end{center}
\end{table}

A remarkable effort has been devoted to the search for LFV in
processes involving charged leptons, resulting in impressive bounds in
some channels. This has been particularly well motivated after the
discovery of neutrino flavor oscillations, which imply that charged
lepton flavor violating processes must exist, although perhaps with
low rates.

In what concerns HLFV, the ATLAS and CMS collaborations have searched
for $h \to \ell_i \ell_j$, setting limits for the corresponding
branching ratios in the $10^{-5}-10^{-3}$ ballpark, as shown in
Tab.~\ref{tab:HLFVsignals}.~\footnote{We define $\BR(h \to \ell_i
  \ell_j) = \BR(h \to \ell_i^+ \ell_j^-) + \BR(h \to \ell_i^-
  \ell_j^+)$.} The CMS limit on $\BR(h \to \mu e)$ has been obtained
using $\sqrt s=8$ TeV data, whereas the rest of ATLAS and CMS limits
have been updated including $\sqrt s=13$ TeV data. Dedicated
strategies can in principle improve these limits substantially with
future LHC data~\cite{Harnik:2012pb}, in particular in the 14 TeV
HL-LHC phase \cite{Banerjee:2016foh}. Furthermore, HLFV can also be
searched for at $e^+ e^-$ colliders, which offer a very clean
environment, perfectly suited for the exploration of the Higgs boson
properties. As shown by several analyses
\cite{Banerjee:2016foh,Chakraborty:2016gff,Chakraborty:2017tyb,Qin:2017aju},
the planned future $e^+ e^-$ facilities (CEPC, FCC-ee and ILC) can
probe HLFV branching ratios as low as $10^{-5}-10^{-4}$, improving on
the current LHC limits by about one order of magnitude for channels
involving the $\tau$ lepton.

\begin{table}[tb!]
\centering
\begin{tabular}{ccc}
\toprule
LFV Process BR & Present Bound & Future Sensitivity  \\
\midrule
    $\mu \to  e \gamma$ & $4.2\times 10^{-13}$~\cite{TheMEG:2016wtm}  & $6\times 10^{-14}$~\cite{Baldini:2013ke} \\
    $\tau \to e \gamma$ & $3.3 \times 10^{-8}$~\cite{Aubert:2009ag}& $ \sim3\times10^{-9}$~\cite{Aushev:2010bq}\\
    $\tau \to \mu \gamma$ & $4.4 \times 10^{-8}$~\cite{Aubert:2009ag}& $ \sim3\times10^{-9}$~\cite{Aushev:2010bq} \\
$\mu \rightarrow e e e$ &  $1.0 \times 10^{-12}$~\cite{Bellgardt:1987du} &  $\sim10^{-16}$~\cite{Blondel:2013ia} \\
    $\tau \rightarrow e e e$ & $2.7\times10^{-8}$~\cite{Hayasaka:2010np} &  $\sim 10^{-9}$~\cite{Aushev:2010bq} \\
    $\tau \rightarrow \mu \mu \mu$ & $2.1\times10^{-8}$~\cite{Hayasaka:2010np} & $\sim 10^{-9}$~\cite{Aushev:2010bq} \\
    $\tau^- \rightarrow e^- \mu^+ \mu^-$ &  $2.7\times10^{-8}$~\cite{Hayasaka:2010np} & $\sim 10^{-9}$~\cite{Aushev:2010bq} \\
    $\tau^- \rightarrow \mu^- e^+ e^-$ &  $1.8\times10^{-8}$~\cite{Hayasaka:2010np} & $\sim 10^{-9}$~\cite{Aushev:2010bq} \\    
$\tau^- \rightarrow e^+ \mu^- \mu^-$ &  $1.7\times10^{-8}$~\cite{Hayasaka:2010np} & $\sim 10^{-9}$~\cite{Aushev:2010bq} \\
$\tau^- \rightarrow \mu^+ e^- e^-$ &  $1.5\times10^{-8}$~\cite{Hayasaka:2010np} & $\sim 10^{-9}$~\cite{Aushev:2010bq} \\
$\mu^-, \mathrm{Ti} \rightarrow e^-, \mathrm{Ti}$ &  $4.3\times 10^{-12}$~\cite{Dohmen:1993mp} & $\sim10^{-18}$~\cite{PRIME} \\
    $\mu^-, \mathrm{Au} \rightarrow e^-, \mathrm{Au}$ & $7\times 10^{-13}$~\cite{Bertl:2006up} & \\
    $\mu^-, \mathrm{Al} \rightarrow e^-, \mathrm{Al}$ &  & $10^{-15}-10^{-18}$~\cite{Pezzullo:2017iqq} \\
    $\mu^-, \mathrm{SiC} \rightarrow e^-, \mathrm{SiC}$ &  & $10^{-14}$~\cite{Natori:2014yba} \\
\bottomrule
\end{tabular}
\caption{Current experimental bounds and future sensitivities for several LFV observables of interest.
\label{tab:lfv}
}
\end{table}

The NP degrees of freedom and interactions leading to $h \to \ell_i
\ell_j$ also generate other LFV processes, such as $\ell_i \to \ell_j
\gamma$. Since these are subject to much stronger experimental bounds,
they tend to be crucial constraints in most specific models and must
be considered in any phenomenological study on HLFV
decays. Tab. \ref{tab:lfv} collects the current bounds and future
sensitivities for several LFV processes of interest. Muon LFV
observables have the best experimental limits due to the existence of
high-intensity muon beams, while the branching ratios for tau LFV
decays are bound to be below $\sim 10^{-8}$. The most constraining
processes in many models are the $\ell_i \to \ell_j \gamma$ radiative
decays. The MEG collaboration has established the strong limit
$\BR(\mu \to e \gamma) < 4.2 \cdot 10^{-13}$, a bound that will be
improved by about an order of magnitude in the MEG-II
phase~\cite{Baldini:2013ke}. Regarding the $\ell_i \to \ell_j \ell_k
\ell_k$ 3-body decays, the $\mu \to e e e$ branching ratio sensitivity
is expected to be improved by four orders of magnitude by the Mu3e
experiment~\cite{Blondel:2013ia}. Finally, the most spectacular
progress in the search for LFV are expected in $\mu-e$ conversion
experiments, which are also expected to improve the current limits for
different nuclei by several orders of
magnitude~\cite{PRIME,Pezzullo:2017iqq,Natori:2014yba}. See
\cite{Calibbi:2017uvl} for an experimental and theoretical review of
the current situation in charged lepton flavor violation experiments.

\section{Model-independent considerations}
\label{sec:eft}

In order to explore HLFV in a model-independent way, it proves
convenient to adopt an approach based on Effectively Field Theory
(EFT). This is particularly well motivated due to lack of NP signals
at the LHC, which arguably implies that any new particles responsible
for HLFV would lie clearly above the electroweak scale. We will now
continue along the lines of
\cite{Herrero-Garcia:2016uab}.~\footnote{See also the comprehensive
  Ref. \cite{Dorsner:2015mja} for a similar reasoning in a multi-Higgs
  EFT that further strengthens the case for potentially large HLFV
  effects in the 2HDM.}

In addition to canonical kinetic terms, the SM Lagrangian contains the
Yukawa terms for the leptons
\begin{equation} \label{eq:SMY}
- \mathcal L_{\rm SM}^Y = \bar \ell \, \Gamma_e \, e \, \varphi + \hc \, ,
\end{equation}
where
\begin{align}
\ell = \left( \begin{array}{c}
\nu \\
e \end{array} \right)_L \sim \left( {\bf 1}, {\bf 2}, -\frac{1}{2} \right) \, , \quad
e \equiv e_R \sim \left( {\bf 1}, {\bf 1}, -1 \right) \quad \text{and} \quad
\varphi = \left( \begin{array}{c}
\varphi^+ \\
\varphi^0 \end{array} \right) \sim \left( {\bf 1},{\bf 2}, \frac{1}{2} \right) \nonumber
\end{align}
denote the SM lepton $\rm SU(2)_L$ doublets and singlets and Higgs
doublet, respectively, and we give the representation under the SM
gauge group $\rm SU(3)_c \times SU(2)_L \times U(1)_Y$. $\Gamma_e$ is
a $3 \times 3$ complex matrix that can be taken to be diagonal without
loss of generality. Therefore, the three lepton flavors are exactly
conserved in the SM, which then possesses a $G_f = \rm{U(1)_e \times
  U(1)_\mu \times U(1)_\tau}$ global flavor symmetry.~\footnote{The
  global flavor symmetry of the SM is known to be broken since the
  experimental observation of neutrino flavor oscillations. Therefore,
  the NP behind the generation of neutrino masses must necessarily
  violate $G_f$ and induce LFV processes such as $\ell_i \to \ell_j
  \gamma$ and $h \to \ell_i \ell_j$. The resulting rates in some
  specific models are too low to be observed in any foreseeable
  experiment. For instance, in the SM minimally extended with Dirac
  neutrino masses one expects tiny LFV branching ratios, as low as
  $\BR(\mu \to e \gamma) \sim 10^{-55}$~\cite{Petcov:1976ff} or $\BR(h
  \to \tau \mu) \sim 10^{-56}$~\cite{Herrero-Garcia:2016uab}. However,
  this is not a generic expectation, since these rates get hugely
  enhanced in most NP scenarios. We refer to Sec.~\ref{sec:numass} for
  more details about the connection between HLFV and neutrino masses.}

The flavor symmetry $G_f$ gets generally broken in the presence of
NP. This can be generically parametrized by means of
non-renormalizable gauge-invariant operators of dimension $d>4$ that
encode the LFV effects induced by unknown heavy states,
\begin{equation}
\mathcal L_{\rm EFT} = \frac{C_i}{\Lambda^{d-4}} \, Q_i + \hc \, .
\end{equation}
Here $\Lambda$ is the scale of NP, of the order of the masses of the
states whose decoupling induces the dimension-$d$ operator $Q_i$, and
$C_i$ the associated Wilson coefficient. There are many of such LFV
operators. However, the only dimension-six operator giving rise to
Higgs LFV is $Q_{e \varphi}$, defined as
\begin{equation} \label{eq:QeH}
Q_{e \varphi} = \left( \varphi^\dagger \varphi\right) \, \left( \bar \ell \, e \, \varphi \right) \, .
\end{equation}
This operator was first highlighted in the context of HLFV in
\cite{Harnik:2012pb}, later denoted \textit{the Yukawa operator} in
\cite{Herrero-Garcia:2016uab} and is one of the operators in the
Warsaw basis of the Standard Model Effective Field
Theory~\cite{Grzadkowski:2010es}.~\footnote{In what concerns effective
  operators, we follow the notation of {\tt
    DsixTools}~\cite{Celis:2017hod}.} Any additional gauge-invariant
dimension-six operator leading to HLFV can be shown to be redundant,
and therefore reducible to $Q_{e \varphi}$ by using equations of
motion, Fierz transformations or other field redefinitions
\cite{Harnik:2012pb}. Therefore, all HLFV effects are encoded (at
least in scenarios leading to NP contributions of dimension six) by
$Q_{e \varphi}$.

After electroweak symmetry breaking, the SM Yukawa term in
Eq. \eqref{eq:SMY} and the NP contribution encoded in $Q_{e \varphi}$
add up to
\begin{align}
\mathcal L_{\rm SM}^Y + \mathcal L_{\rm EFT} &\supset - \, \bar e_L \left[ \frac{v}{\sqrt{2}} \, \left( \Gamma_e - C_{e \varphi} \, \frac{v^2}{2 \Lambda^2} \right) + \frac{h}{\sqrt{2}} \, \left( \Gamma_e - 3 \, C_{e \varphi} \, \frac{v^2}{2 \Lambda^2} \right) \right] e_R + \hc \nonumber \\
&= - \, \bar e_L \left[ \mathcal{M}_e + h \, \mathcal{Y}_e \right] e_R + \hc \, ,
\end{align}
where
\begin{equation}
\mathcal{M}_e = \frac{v}{\sqrt{2}} \, \left( \Gamma_e - C_{e \varphi} \, \frac{v^2}{2 \Lambda^2} \right)
\end{equation}
is the $3 \times 3$ charged lepton mass matrix and
\begin{equation}
\mathcal{Y}_e = \frac{1}{\sqrt{2}} \, \left( \Gamma_e - 3 \, C_{e \varphi} \, \frac{v^2}{2 \Lambda^2} \right)
\end{equation}
are the Higgs boson couplings to a pair of charged leptons. We have
used the decomposition $\varphi^0 = \frac{1}{\sqrt{2}} (h + i \, G^0 +
v)$, with $h$ the physical Higgs boson with a mass $m_h \simeq 125$
GeV, $G^0$ the Goldstone boson that constitutes the longitudinal
component of the massive $Z$-boson and $v \simeq 246$ GeV the
electroweak vacuum expectation value (VEV). It is clear that, in
general, the matrices $\mathcal{M}_e$ and $\mathcal{Y}_e$ are not
diagonal in the same basis. In fact, in the mass basis, defined by
\begin{equation}
V_{e_L}^\dagger \mathcal{M}_e V_{e_R} = \widehat{\mathcal{M}}_e = \text{diag} (m_e , m_\mu , m_\tau) \, ,
\end{equation}
the Higgs boson couplings to charged leptons read
\begin{equation} \label{eq:yEFT}
g_{h \ell \ell} = V_{e_L}^\dagger \mathcal{Y}_e V_{e_R} = \frac{1}{v} \, \widehat{\mathcal{M}}_e - \frac{v^2}{\sqrt{2} \Lambda^2} \, V_{e_L}^\dagger C_{e \varphi} V_{e_R}  \, .
\end{equation}
While the first term in Eq. \eqref{eq:yEFT} is proportional to the
charged lepton masses, the second one can in general contain
off-diagonal entries and induce HLFV processes. In particular, this
piece leads to the $h \to \ell_i \ell_j$ decays, with $i \ne j$. One
finds
\begin{equation}
\BR (h \to \ell_i \ell_j) = \frac{m_h}{8 \pi \Gamma_h} \, \left( |g_{h \ell_i \ell_j}|^2 + |g_{h \ell_j \ell_i}|^2 \right) \, ,
\end{equation}
where $\Gamma_h$ ($\simeq 4$ MeV in the SM) is the total Higgs boson
decay width.

So far, we only discussed the $Q_{e \varphi}$ operator, which induces
the HLFV decays $h \to \ell_i \ell_j$ we are interested in. However,
in a complete ultraviolet theory other operators will be generated as
well. In particular, operators that give rise to other LFV processes,
such as $\ell_i \to \ell_j \gamma$, with much stronger experimental
bounds. At dimension six, two gauge-invariant operators of this type
exist~\cite{Grzadkowski:2010es},
\begin{equation}
Q_{eW} = \left( \bar \ell \sigma^{\mu \nu} e \right) \tau^I \varphi \, W^I_{\mu \nu} \quad \text{and} \quad Q_{eB} = \left( \bar \ell \sigma^{\mu \nu} e \right) \varphi \, B_{\mu \nu} \, ,
\end{equation}
where $\tau^I$, with $I = 1,2,3$, are the Pauli matrices. At low
energies, these two operators are matched to the dimension-five
photonic dipole operator~\footnote{Explicit expressions for the
  tree-level matching can be found in \cite{Jenkins:2017jig}.}
\begin{equation} \label{eq:OpDip}
\mathcal{O}_{e\gamma} = \bar e_L \sigma^{\mu \nu} e_R  \, F_{\mu \nu} \, ,
\end{equation}
which is directly responsible for the $\ell_i \to \ell_j \gamma$
radiative LFV decays. Defining the contribution of
$\mathcal{O}_{e\gamma}$ to the low-energy effective Lagrangian as
\begin{equation} \label{eq:WCDip}
\mathcal{L}_{\rm EFT}^{\rm low} = \frac{L_{e \gamma}}{v} \, \mathcal{O}_{e\gamma} + \hc \, ,
\end{equation}
where $L_{e \gamma}$ is its Wilson coefficient, the resulting
branching ratios are
\begin{equation}
\BR(\ell_i \to \ell_j \gamma) = \frac{m_i^3}{4 \pi v^2 \, \Gamma_i} \left( |\left(L_{e \gamma} \right)_{ij}|^2 + |\left(L_{e \gamma} \right)_{ji}|^2 \right) \, ,
\end{equation}
with $m_i$ and $\Gamma_i$ the mass and total decay width of the
charged lepton $\ell_i$, respectively.

Any theory that induces $Q_{e \varphi}$ will also generate
$\mathcal{O}_{e \gamma}$, since these operators transform in the same
way under flavor and chiral symmetries~\cite{Dorsner:2015mja} and mix
under renormalization group
evolution~\cite{Alonso:2013hga}. Therefore, one cannot simply get rid
of the latter. However, different NP scenarios predict a different
balance between the $C_{e \varphi}$ and $L_{e \gamma}$ Wilson
coefficients, and this is what determines the magnitude of the allowed
HLFV effects in a specific model. Let us consider an example to
illustrate this connection: a model inducing predominantly $\left(Q_{e
  B}\right)_{12}$ at the high-energy scale $\Lambda$. In this case the
operator $\left(Q_{e \varphi}\right)_{12}$ gets induced due to
renormalization group running while $\left(\mathcal{O}_{e
  \gamma}\right)_{12}$ is obtained after matching at the electroweak
scale. Since the HLFV operator is induced by operator mixing effects,
the resulting coefficient is suppressed and one expects the relation
$\BR(h \to \mu e) \simeq 10^{-14} \, \log^2 \left( m_h / \Lambda
\right) \, \BR(\mu \to e \gamma)$, which clearly precludes the
observation of the HLFV decay. More generally, in models with $L_{e
  \gamma} \sim C_{e \varphi}$ or $L_{e \gamma} > C_{e \varphi}$, as in
the example we just considered, the strong constraints derived from
the non-observation of $\ell_i \to \ell_j \gamma$ would imply tiny
HLFV rates. In contrast, models predicting $L_{e \gamma} \ll C_{e
  \varphi}$ may accommodate sizable HLFV effects. As we will see in
Sec. \ref{subsec:EFT-2HDM}, the 2HDM is one of such models.

\section{HLFV in the 2HDM}
\label{sec:2HDM}

We now concentrate on the 2HDM. First, in Sec. \ref{subsec:EFT-2HDM}
we particularize the previous model-independent discussion to the case
of a 2HDM in order to motivate this model as the perfect scenario to
obtain large HLFV rates. Sec. \ref{subsec:notation-2HDM} will
introduce our notation and conventions for the 2HDM. Finally, we will
concentrate on $\tau \mu$ flavor violation, discuss $\tau \to \mu
\gamma$ in the 2HDM in Sec.~\ref{subsec:taumugamma} and show some
selected phenomenological results on $h \to \tau \mu$ in the 2HDM in
Sec.~\ref{subsec:pheno-2HDM}.

\subsection{EFT motivation}
\label{subsec:EFT-2HDM}

\begin{figure}
\centering
\includegraphics[width=0.4\textwidth]{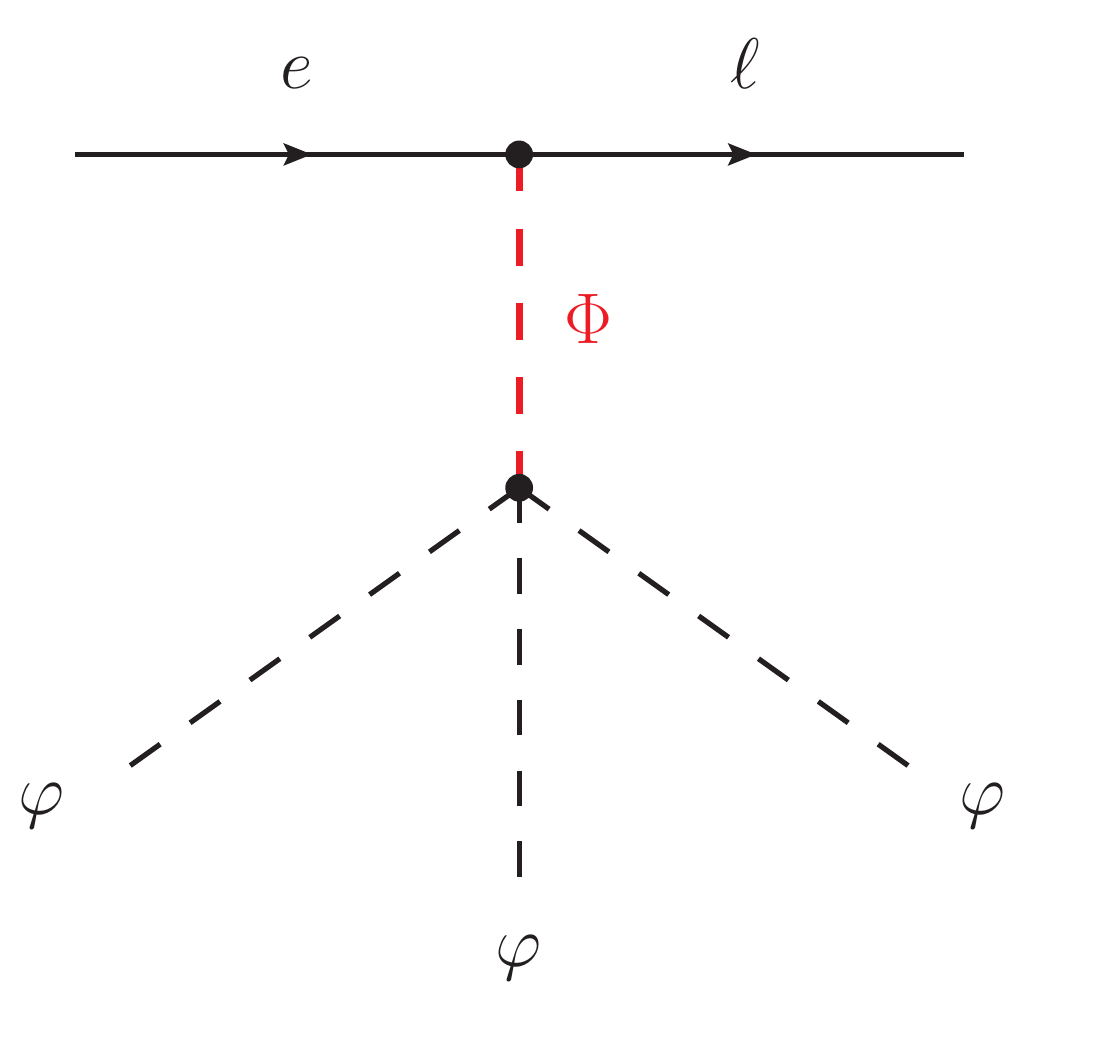}
\caption{2HDM ultraviolet completion of the $Q_{e \varphi}$
  operator. Here $\Phi$ is a heavy new scalar $\rm SU(2)_L$
  doublet. See \cite{Herrero-Garcia:2016uab} for details.}
\label{fig:topo}
\end{figure}

A NP model with a second scalar $\rm SU(2)_L$ doublet would induce the
$Q_{e \varphi}$ operator, as shown in Fig. \ref{fig:topo}. This is one
of the possible topologies contributing to the HLFV operator, as
discussed in great detail in \cite{Herrero-Garcia:2016uab}. The scalar
$\Phi$ must be an $\rm SU(2)_L$ doublet for the diagram to be gauge
invariant, and would be identified with a second Higgs doublet in a
2HDM. Moreover, it should be noticed that this topology requires
\textit{both} scalar doublets to couple to leptons. The $\Phi$
coupling is clearly shown, and $\varphi$ was already assumed to couple
to leptons, see Eq. \eqref{eq:SMY}. For this reason, the 2HDM behind
the generation of this topology would be a \textit{type-III} 2HDM, in
which both Higgs doublets are allowed to couple to leptons in a
general way.~\footnote{This excludes more popular versions of the
  2HDM. In particular it excludes 2HDMs with natural flavor
  conservation \cite{Glashow:1976nt,Paschos:1976ay}, like the type-II
  2HDM included in supersymmetric models. We refer to
  \cite{Branco:2011iw} for a comprehensive review of the 2HDM and all
  its variants.}

We note that $Q_{e \varphi}$ is generated at tree-level in this
scenario, thus enhancing the Wilson coefficient $C_{e \varphi}$. We
expect
\begin{equation}
\frac{C_{e \varphi}}{\Lambda^2} \sim \frac{\lambda \, y_\Phi}{m_\Phi^2} \, ,
\end{equation}
where $\lambda$ and $y_\Phi$ are the quartic scalar and Yukawa
couplings involved in the topology shown in Fig. \ref{fig:topo} and
$m_\Phi$ the $\Phi$ scalar mass term. Next, we consider the generation
of $\mathcal{O}_{e \gamma}$. This operator can be generated by
attaching one of the external $\varphi$ lines in Fig. \ref{fig:topo}
to a charged lepton line and then adding the photon. There are,
however, three effects that suppress the generation of this operator
in the 2HDM:
\begin{itemize}
\item The $\mathcal{O}_{e \gamma}$ operator gets induced at the 1-loop
  level.
\item Closing the loop by attaching the $\varphi$ line to the charged
  lepton line introduces a charged lepton mass insertion.
\item The $\mathcal{O}_{e \gamma}$ operator requires a chirality flip,
  which in a SM extension with only new scalar fields (such as the
  2HDM) implies an additional charged lepton mass insertion.
\end{itemize}
These considerations allow us to estimate
\begin{equation}
\left( L_{e \gamma} \right)_{ij} \sim \left( \frac{m_i}{v} \right)^2
\, \frac{1}{16 \, \pi^2} \, \left( C_{e \varphi} \right)_{ij} \ll
\left( C_{e \varphi} \right)_{ij} \, .
\end{equation}
This estimate is known to be very poor due to the presence of several
additional contributions in a complete 2HDM, as we will show in
Sec. \ref{subsec:pheno-2HDM}. Nevertheless, it serves as a good
motivation to consider this scenario as potentially promising in what
concerns HLFV, since the required hierarchy between $C_{e \varphi}$
and $L_{e \gamma}$ is naturally obtained.

\subsection{2HDM: model basics}
\label{subsec:notation-2HDM}

In the following, we consider the general 2HDM, usually referred to as
type-III 2HDM, and denote the two Higgs doublets as $\varphi_1$ and
$\varphi_2$. In contrast to other variants of the 2HDM, no distinction
between the two Higgs doublets is introduced. This has two
consequences for our discussion. First, both $\varphi_1$ and
$\varphi_2$ are allowed to couple to all fermion species, and in
particular to leptons, a fundamental ingredient for the generation of
HLFV effects, see Sec. \ref{subsec:EFT-2HDM}. And second, one can
perform arbitrary $\rm U(2)$ basis transformations in $\{ \varphi_1 ,
\varphi_2 \}$ space, without any impact on physical observables. This
freedom can be used to go to a specific basis in which only one Higgs
doublet acquires a VEV, the so-called Higgs basis
\cite{Georgi:1978ri,Donoghue:1978cj,Botella:1994cs}. In this basis,
the scalar potential of the model is given by~\footnote{We follow the
  conventions of \cite{Davidson:2005cw}, with a slightly different
  notation.}
\begin{align}
\mathcal{V} &= \, M_{11}^2 \, \varphi_1^\dagger \varphi_1 + M_{22}^2 \, \varphi_2^\dagger \varphi_2 - \left( M_{12}^2 \, \varphi_1^\dagger \varphi_2 + \hc \right) \nonumber \\
& + \frac{\Lambda_1}{2} \left(\varphi_1^\dagger \varphi_1\right)^2 + \frac{\Lambda_2}{2} \left(\varphi_2^\dagger \varphi_2\right)^2 + \Lambda_3 \left(\varphi_1^\dagger \varphi_1\right)\left(\varphi_2^\dagger \varphi_2\right) + \Lambda_4 \left(\varphi_1^\dagger \varphi_2\right)\left(\varphi_2^\dagger \varphi_1\right) \nonumber \\
&+ \left[ \frac{\Lambda_5}{2} \left(\varphi_1^\dagger \varphi_2\right)^2 + \Lambda_6 \left(\varphi_1^\dagger \varphi_1\right)\left(\varphi_1^\dagger \varphi_2\right) + \Lambda_7 \left(\varphi_2^\dagger \varphi_2\right)\left(\varphi_1^\dagger \varphi_2\right) + \hc \right] \, ,
\end{align}
and the Higgs doublets can be decomposed as
\begin{equation} \label{h1h2def}
\varphi_1 = \left( \begin{array}{c}
G^+ \\
\frac{1}{\sqrt{2}} \left( v + \phi_1^0 + i \, G^0 \right)
\end{array} \right) \quad , \quad \varphi_2 = \left( \begin{array}{c}
H^+ \\
\frac{1}{\sqrt{2}} \left( \phi_2^0 + i \, A \right)
\end{array} \right) \, .
\end{equation}
Here $\phi_1^0$, $\phi_2^0$ and $A$ are neutral scalars, $H^+$ is a
charged scalar and $G^+$ and $G^0$ are Goldstone bosons. Assuming CP
conservation, the CP-even states $\phi_1^0$ and $\phi_2^0$ do not mix
with the CP-odd state $A$, which is a physical mass
eigenstate. $\phi_1^0$ and $\phi_2^0$ are related to the mass
eigenstates $h$ and $H$ (with $m_h < m_H$) as
\begin{equation}
\left( \begin{array}{c}
h \\
H \end{array} \right) = \left( \begin{array}{cc}
s_{\beta - \alpha} & c_{\beta - \alpha} \\
c_{\beta - \alpha} & - s_{\beta - \alpha} \end{array} \right) \, \left( \begin{array}{c}
\phi_1^0 \\
\phi_2^0 \end{array} \right) \, , \label{hHrot}
\end{equation}
where $s_{\beta - \alpha} \equiv \sin (\beta - \alpha)$, $c_{\beta -
  \alpha} \equiv \cos (\beta - \alpha)$ and $\beta - \alpha$ is a
physical mixing angle. The lighest CP-even state, $h$, is identified
with the Higgs boson discovered at the LHC. With these definitions at
hand, one can derive several relations between the potential
parameters and the physical Higgs masses~\cite{Davidson:2005cw},
\begin{eqnarray}
m_{H^+}^2 &=& M_{22}^2 + \frac{v^2}{2} \Lambda_3 \, , \label{eqfirst} \\
m_A^2 - m_{H^+}^2 &=& - \frac{v^2}{2} \left( \Lambda_5 - \Lambda_4 \right) \, , \\
m_H^2 + m_h^2 - m_A^2 &=& v^2 \left( \Lambda_1 + \Lambda_5 \right) \, , \\
(m_H^2 - m_h^2)^2 &=& \left[ m_A^2 + \left( \Lambda_5 - \Lambda_1 \right) v^2 \right]^2 + 4 \, \Lambda_6^2 \, v^4 \, , \\
\sin \left[ 2(\beta - \alpha) \right] &=& - \frac{2 \, \Lambda_6 \, v^2}{m_H^2 - m_h^2} \, . \label{eqlast}
\end{eqnarray}

Let us now discuss the Yukawa interactions of the model. Again, we use
the freedom to choose specific weak bases. In the Higgs basis for the
scalar doublets and the mass basis for the fermions, the Yukawa
Lagrangian can be written as
\begin{align}
- \mathcal L_{\rm 2HDM}^Y =& \, \frac{\sqrt{2}}{v} \, \left( \bar q \, K^\ast \, \widehat{\mathcal{M}}_u  \, u \, \widetilde{\varphi}_1 + \bar q \, \widehat{\mathcal{M}}_d  \, d \, \varphi_1 + \bar \ell \, \widehat{\mathcal{M}}_e  \, e \, \varphi_1 \right) \nonumber \\
& + \bar q \, \rho_u \, u \, \widetilde{\varphi}_2 + \bar q \, \rho_d \, d \, \varphi_2 + \bar \ell \, \rho_e \, e \, \varphi_2 + \hc \, . \label{eq:Y2HDM}
\end{align}
Here we denote $\widetilde{\varphi}_a = i \, \sigma_2 \,
\varphi_a^\ast$, with $a = 1,2$, and define the diagonal matrices
$\widehat{\mathcal{M}}_u = \text{diag}(m_u,m_c,m_t)$ and
$\widehat{\mathcal{M}}_d = \text{diag}(m_d,m_s,m_b)$. $K$ is the CKM
matrix and $\rho_{u,d,e}$ are general $3 \times 3$ complex matrices in
flavor space, which in the following will be assumed to be Hermitian
for simplicity. Using Eqs. \eqref{h1h2def} and \eqref{hHrot}, and
expanding in $\rm SU(2)_L$ indices, the leptonic part of
Eq. \eqref{eq:Y2HDM} can be rewritten as
\begin{align}
- \mathcal L_{\rm 2HDM}^Y \supset& \, \, \bar e_L \left( \frac{1}{v} \, \widehat{\mathcal{M}}_e \, s_{\beta - \alpha} + \frac{1}{\sqrt{2}} \, \rho_e \, c_{\beta - \alpha} \right) e_R \, h \nonumber \\
& + \, \bar e_L \left( \frac{1}{v} \, \widehat{\mathcal{M}}_e \,  c_{\beta - \alpha} - \frac{1}{\sqrt{2}} \, \rho_e \, s_{\beta - \alpha} \right) e_R \, H \nonumber \\
& + \, \frac{i}{\sqrt{2}} \, \bar e_L \, \rho_e \, e_R \, A + \bar \nu_L \left( U^\dagger \rho_e \right) \, e_R \, H^+ + \hc \, , \label{eqYuk}
\end{align}
where $U$ is the PMNS matrix. This expression allows us to extract the
couplings of the neutral scalars of the model to leptons. By following
the same steps with the quarks, one finds the general expressions
\begin{align}
g_{hff^\prime} &= \frac{1}{v} \, \widehat{\mathcal{M}}_f \, s_{\beta - \alpha} + \frac{1}{\sqrt{2}} \, \rho_f \, c_{\beta - \alpha} \, , \label{eq:g2HDM1} \\
g_{Hff^\prime} &= \frac{1}{v} \, \widehat{\mathcal{M}}_f \, c_{\beta - \alpha} - \frac{1}{\sqrt{2}} \, \rho_f \, s_{\beta - \alpha} \, , \label{eq:g2HDM2} \\
g_{Aff^\prime} &= \frac{i}{\sqrt{2}} \, s_f \, \rho_f \, , \label{eq:g2HDM3}
\end{align}
where $f=u,d,e$. He have introduced $s_f = +1$ for down-type quarks
and charged leptons and $s_f = -1$ for up-type
quarks. Eq. \eqref{eq:g2HDM1} must be compared to the general
expression in Eq. \eqref{eq:yEFT}. Again, the first term is diagonal,
whereas the second may contain off-diagonal entries. We therefore
conclude that the $\rho_e$ matrix is the source of the HLFV processes
that we are about to discuss.

Finally, the neutral scalar couplings to a pair of gauge bosons are
fully dictated by the gauge symmetry. One has
\begin{align}
C_{hWW} &= s_{\beta - \alpha} \, C_{hWW}^{\text{SM}} \, , \label{eq:g2HDM4} \\
C_{HWW} &= c_{\beta - \alpha} \, C_{hWW}^{\text{SM}} \, , \label{eq:g2HDM5} \\
C_{AWW} &= 0 \, , \label{eq:g2HDM6}
\end{align}
and the couplings to a pair of $Z$-bosons follow the same
proporcionalities.

\subsection{$\tau \to \mu \gamma$ in the 2HDM}
\label{subsec:taumugamma}

Given the strong experimental bounds on $\mu e$ flavor violating
processes, we will concentrate on $\tau$ LFV. In particular, we will
consider $\tau \mu$ LFV, and therefore discuss $h \to \tau \mu$ and
the related $\tau \to \mu \gamma$. The $h \to \tau \mu$ HLFV decay is
the main focus of this manuscript and we show some phemenological
results in Sec. \ref{subsec:pheno-2HDM}. However, in order to assess
the observability of this process, one must take into account the
strong constraint coming from $\BR(\tau \to \mu \gamma)$, which we now
proceed to evaluate in the 2HDM. \\

\begin{figure}
\centering
\includegraphics[width=0.45\textwidth]{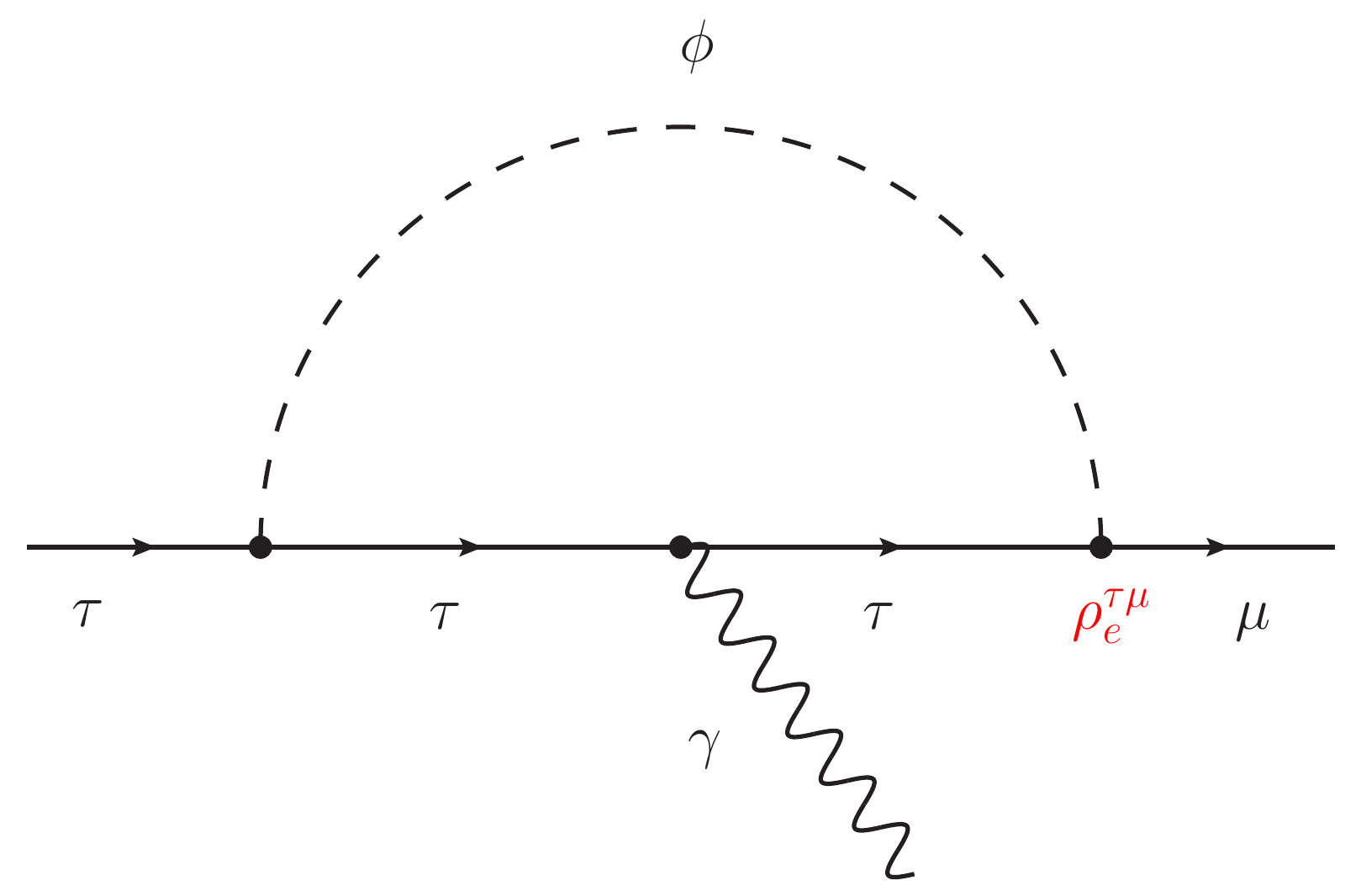}
\includegraphics[width=0.45\textwidth]{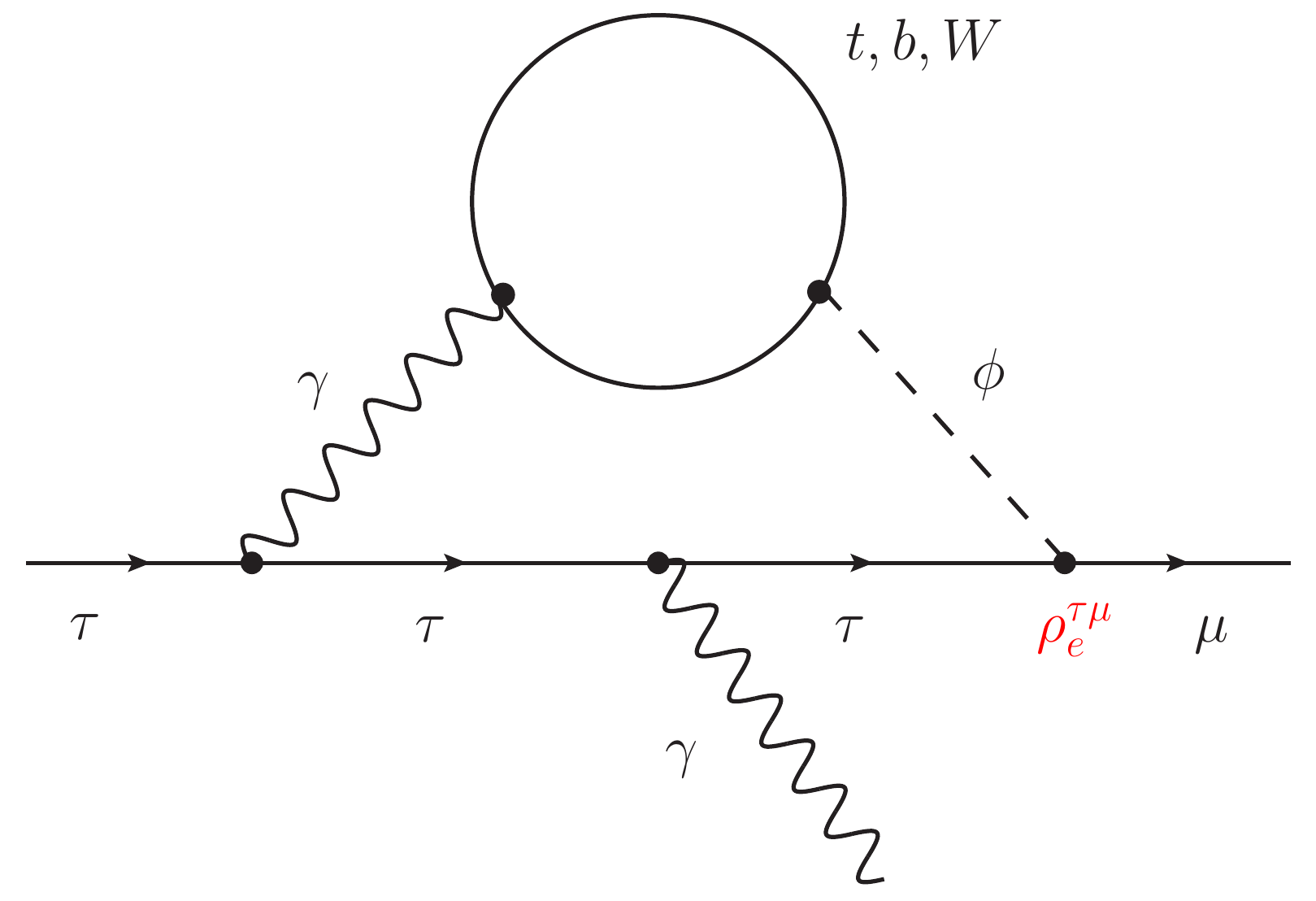}
\caption{Most important contributions to $\tau \to \mu \gamma$ in the
  2HDM. Here $\phi = h, H, A$. To the left, 1-loop diagrams with
  neutral Higgs bosons and charged leptons in the loop. To the right,
  2-loop Barr-Zee diagrams with an internal photon and a third
  generation quark or a $W$-boson. The LFV vertex proportional to
  $\rho_e^{\tau \mu}$ is explicitly indicated.}
\label{fig:loops}
\end{figure}

The $\tau \to \mu \gamma$ radiative decay is induced by the dipole
operator defined by Eqs. \eqref{eq:OpDip} and \eqref{eq:WCDip}. It
proves convenient to define the form factors $A_L$ and $A_R$ as
\begin{equation}
\left( L_{e \gamma} \right)_{\tau \mu} = \frac{e \, m_\tau v}{2} \, A_R^{\tau \mu}
  \quad , \quad 
\left( L_{e \gamma} \right)_{\mu \tau}^\ast = \frac{e \, m_\tau v}{2} \, A_L^{\tau \mu} \, .
\end{equation}
Since we assume the matrix $\rho$ to be Hermitian, $|g_{h \tau \mu}| =
|g_{h \mu \tau}|$, which implies $|A_L| = |A_R| \equiv |A|$. We just
need to determine the most relevant contributions to the form factor
$A$ in the 2HDM.

It is well known that in the 2HDM 2-loop contributions to $\ell_i \to
\ell_j \gamma$ may easily dominate over 1-loop
ones~\cite{Bjorken:1977br}. The reason is easy to understand. Dipole
transitions require a chirality flip. In a 1–loop diagram with a
virtual scalar in the loop, two chirality flips take place in the
Yukawa vertices, and therefore one more is required in the fermion
propagator, giving a total of three. This largely suppresses the loop
amplitude, which explains why 2-loop diagrams with only one chirality
flip can be dominant even if one pays the extra loop suppression
factor of $1/(16 \pi^2)$. In particular, 2-loop Barr-Zee diagrams
\cite{Barr:1990vd} can easily dominate if the involved scalar fields
have large couplings to the virtual fermions or bosons running in the
loops. Taking all these ingredients into account, the authors of
Ref. \cite{Davidson:2010xv} identified three main contributions to
$\tau \to \mu \gamma$ in the type-III 2HDM:
\begin{itemize}
\item 1-loop diagrams with neutral Higgs bosons and charged leptons in
  the loop
\item 2-loop Barr-Zee diagrams with an internal photon and a third
  generation quark
\item 2-loop Barr-Zee diagrams with an internal photon and a $W$-boson
\end{itemize}
These contributions were computed in \cite{Chang:1993kw} and are shown
in Fig. \ref{fig:loops}. One can write
\begin{equation} \label{eq:contri}
A = \frac{1}{16 \pi^2} \left( A_1 + A_2^{t,b} + A_2^W \right) \, ,
\end{equation}
where the different contributions were computed by
\cite{Davidson:2010xv} and are given explicitly in
Appendix~\ref{app:contri}. Armed with these expressions we are ready
to explore the HLFV phenomenology of the type-III 2HDM.

\subsection{HLFV phenomenology}
\label{subsec:pheno-2HDM}

Following Ref. \cite{Sierra:2014nqa}, we show now some
phenomenological results on HLFV in the type-III 2HDM. We refer to
\cite{Crivellin:2015mga,Omura:2015nja,Crivellin:2015hha,Botella:2015hoa,Liu:2015oaa,Benbrik:2015evd,Bizot:2015qqo,Sher:2016rhh,Herrero-Garcia:2016uab,Wang:2016rvz,Tobe:2016qhz,Qin:2017aju,Babu:2018uik}
for additional HLFV phenomenological studies in the 2HDM. \\

First, we must make an observation about the type-III 2HDM. As already
explained, in this version of the 2HDM one can apply rotations in
Higgs space that modify the Higgs VEVs. For this reason, the usual
ratio of VEVs $\tan \beta$ is not uniquely defined. Given that we are
mainly interested in tau flavor violation, we
define~\cite{Davidson:2005cw}
\begin{equation}
\tan \beta_\tau = \frac{- \rho_e^{\tau \tau}}{\sqrt{2} m_\tau/v} \, .
\end{equation}
We note that $\tan \beta_\tau$ is the physical ratio between the tau
Yukawa coupling and $\sqrt{2} m_\tau/v$, which would correspond to the
usual $\tan \beta$ in the Type-II 2HDM.

Let us now discuss our parameter choices. The results presented here
are based on a random scan of the parameter space, taking the
parameter ranges,
\begin{gather}
200 \, \text{GeV} < m_H < 1000 \, \text{GeV} \, , \nonumber \\
400 \, \text{GeV} < m_A < 1000 \, \text{GeV} \, , \nonumber \\
-5 \, \text{GeV} < m_{H^\pm} - m_A < 5 \, \text{GeV}\ , \nonumber \\
0.7 < \sin (\beta - \alpha) < 1.0 \, , \nonumber \\
0.1 < \tan \beta_\tau < 40 \, . \label{eq:parameter-ranges}
\end{gather}
These are based on the following considerations and experimental
constraints:
\begin{itemize}
\item It proves convenient to use as input the scalar masses, rather
  than the scalar potential parameters. These should nevertheless be
  computed to make sure that they never exceed the perturbative limit
  of $4 \pi$. The $\Lambda_1$, $\Lambda_2$, $\Lambda_4$, $\Lambda_5$
  and $\Lambda_6$ parameters can be computed by means of
  Eqs. \eqref{eqfirst} - \eqref{eqlast}. The remaining $\Lambda$
  parameters do not have any impact on the observables studied here,
  but they can be chosen by demanding that the scalar potential is
  bounded from
  below~\cite{Ivanov:2006yq,Ivanov:2007de,Ferreira:2009jb}.
\item A small $m_{H^\pm} - m_A$ mass difference is required by
  electroweak precision data. In particular, larger values could
  potentially lead to a $T$ oblique parameter value outside the $1 \,
  \sigma$ range determined by \cite{Baak:2012kk}, $T \in \left[ -0.03
    , 0.19 \right]$.
\item The lower limit on $\sin (\beta - \alpha)$ is motivated by the
  fact that the observed Higgs boson has SM-like couplings to fermions
  and gauge bosons. Lower values of $\sin (\beta - \alpha)$ could
  potentially induce deviations, see Eqs. \eqref{eq:g2HDM1} and
  \eqref{eq:g2HDM4}, in tension with the experimental results. In our
  analysis we consider the CMS measurements presented in
  \cite{CMS:2014ega} and require that the signal strengths for $h \to
  \tau \bar \tau , b \bar b , W W^\ast , Z Z^\ast$, defined as $\mu =
  \left( \sigma \times \text{BR} \right) / \left( \sigma \times
  \text{BR} \right)_{\text{SM}}$, are within the CMS 1 $\sigma$ ranges
  \cite{CMS:2014ega}. For the determination of the Higgs production
  cross-section we assume gluon fusion.
\item The lower limit on $m_{H^\pm}$ is motivated by flavor physics
  (mainly B physics, see for instance \cite{Crivellin:2013wna}).
\end{itemize}
Finally, our scan also fixes $m_h = 125.1$
GeV~\cite{Tanabashi:2018oca}. In what concerns the Yukawa matrices,
and in order to reduce the number of free parameters, we will consider
specific textures for the $\rho$ matrices. Inspired by the
\emph{Cheng-Sher ansatz} \cite{Cheng:1987rs}, we express $\rho_e$ as
\begin{equation}
\rho_e^{ij} = - \kappa_{ij} \, \tan \beta_\tau \, \sqrt{\frac{2 \, m_i \, m_j}{v^2}} \, .
\end{equation}
By construction, $\kappa_{\tau \tau} = 1$. However, the other entries
of the $\kappa$ matrix are free parameters. In particular,
$\kappa_{\tau \mu} = \kappa_{\mu \tau}^\ast$ is the relevant parameter
giving rise to the $h \to \tau \mu$ and $\tau \to \mu \gamma$
decays. In our random scan we will take $0.1 < |\kappa_{\tau \mu}| <
3.0$. For the quark $\rho$ matrices we assume the usual Type-II
textures
\begin{equation} \label{ansatz}
\rho_d = - \sqrt{2} \, \tan \beta_\tau \, \frac{\widehat{\mathcal{M}}_d}{v} \quad , \quad 
\rho_u = \sqrt{2} \, \cot \beta_\tau \, \frac{K^\dagger \, \widehat{\mathcal{M}}_u}{v} \, .
\end{equation}
This ansatz is particularly convenient since it ensures compatibility
with the (already constraining) experimental bounds on the Higgs boson
couplings to quarks. Furthermore, it can be regarded as a departure
beyond the popular Type-II 2HDM, with the only deviation in the $\tau
\mu$ coupling~\cite{Kanemura:2005hr}.~\footnote{A modified Cheng-Sher
  ansatz was recently proposed in \cite{Babu:2018uik} in order to
  suppress all Higgs-mediated flavor effects.}

\begin{figure}
\centering
\includegraphics[width=0.6\textwidth]{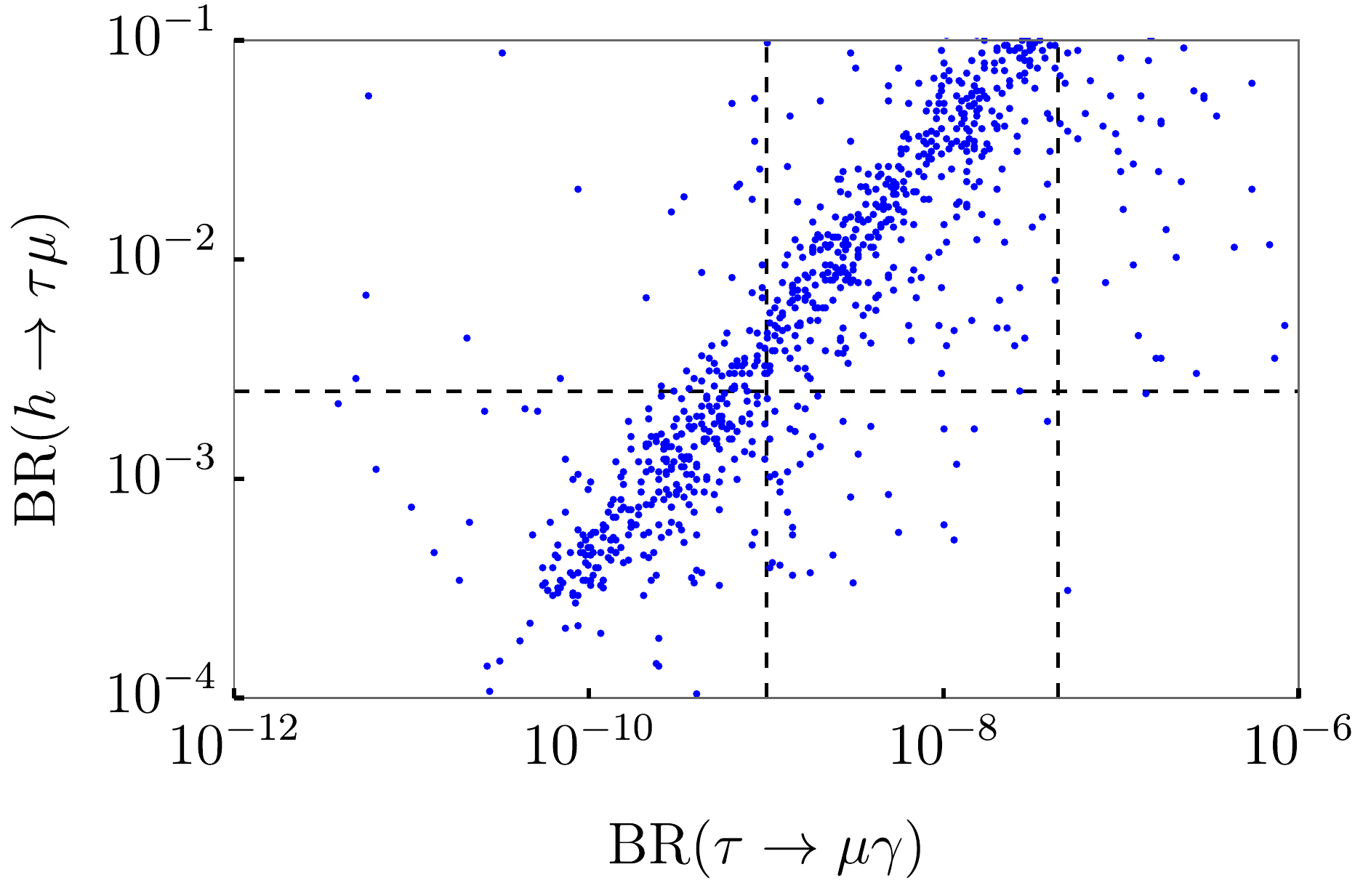}
\caption{$\BR(h \to \tau \mu)$ as a function of $\BR(\tau \to \mu
  \gamma)$ in the type-III 2HDM. The parameters are fixed as explained
  in the text. The vertical lines correspond to the current bound
  $\BR(\tau \to \mu \gamma) < 4.4 \times 10^{-8}$~\cite{Aubert:2009ag}
  and the expected Belle-II sensitivity, estimated to be $\sim
  10^{-9}$~\cite{Aushev:2010bq}. Finally, the horizontal line
  indicates the limit by the CMS collaboration $\BR(h \to \tau \mu) <
  0.0025$~\cite{Sirunyan:2017xzt}.}
\label{fig:pheno}
\end{figure}

After these preliminaries, we are ready to show some results on $h \to
\tau \mu$. Fig.~\ref{fig:pheno} shows $\BR(h \to \tau \mu)$ as a
function of $\BR(\tau \to \mu \gamma)$. The vertical lines shown in
this figure correspond to the current experimental bound $\BR(\tau \to
\mu \gamma) < 4.4 \times 10^{-8}$~\cite{Aubert:2009ag} and the
expected sensitivity of the Belle-II experiment, of about $\sim
10^{-9}$~\cite{Aushev:2010bq}. The horizontal line corresponds to the
limit $\BR(h \to \tau \mu) < 0.0025$, set by the CMS
collaboration~\cite{Sirunyan:2017xzt}. As can be seen from this figure, the
correlation between these two observables is not exact, and this can
be traced back to the different contributions to $\tau \to \mu
\gamma$, which might even cancel in some cases. We find that, in
general, the dominant contribution to the $\tau \to \mu \gamma$
amplitude comes from 2-loop Barr-Zee diagrams with internal $W$
bosons, although the other contributions typically play a relevant
role as well.

The main qualitative message that one can extract from
Fig.~\ref{fig:pheno} is that the type-III 2HDM can induce $h \to \tau
\mu$ rates close to the current bound while being in agreement with
all experimental constraints. All LFV observables increase with
$\kappa_{\tau \mu}$, and in some regions of the parameter space they
can be close to their current experimental limits, explicitly shown in
Fig.~\ref{fig:pheno}. These regions are characterized by $\tan
\beta_\tau \gtrsim 2$, $\sin (\beta - \alpha) \sim 0.9$ and
$\kappa_{\tau \mu} \gtrsim 0.1$.

Finally, some additional comments are in order. A Higgs doublet with
$\mu \tau$ LFV couplings can also address the long-standing muon $g-2$
anomaly. This was studied in relation to the LFV decay $h \to \tau
\mu$ in
\cite{Davidson:2010xv,Crivellin:2013wna,Sierra:2014nqa,Omura:2015nja,Liu:2015oaa,Benbrik:2015evd,Altmannshofer:2016oaq,Wang:2016rvz},
and very recently in \cite{Iguro:2019sly,Wang:2019ngf}. It could also
be linked to the popular $b \to s$
\cite{Crivellin:2015mga,Huang:2015vpt} or $b \to c$
\cite{Crivellin:2015hha} anomalies observed in B-meson decays or be a
crucial ingredient for lepton-flavored electroweak baryogenesis
\cite{Guo:2016ixx}. In fact, the type-III 2HDM with generic Yukawa
couplings has a very rich flavor phenomenology, see
\cite{Crivellin:2013wna}. The analogous quark flavor violating decay
$h \to b s$ was studied in \cite{Crivellin:2017upt}. It is also
remarkable that the 2HDM with a BGL symmetry
\cite{Branco:1996bq,Botella:2009pq,Botella:2011ne} can also lead to
large $h \to \tau \mu$ branching ratios, strongly correlated with
other flavor observables, as shown in \cite{Botella:2015hoa}. Here we
concentrated on the 2HDM. For HLFV studies in other multi-Higgs
doublet models see the interesting works
\cite{Bhattacharyya:2010hp,Bhattacharyya:2012ze,Arroyo:2013tna,Campos:2014zaa}.

\section{2HDMs, neutrino masses and HLFV}
\label{sec:numass}

Neutrino flavor oscillations constitute the only existing experimental
proof of LFV. Since these are sourced by non-zero neutrino masses and
mixings, it is interesting to discuss their possible connection to
HLFV~\cite{Herrero-Garcia:2016uab,Aoki:2016wyl}.

First, one should notice that while the existence of non-zero neutrino
masses and mixings implies the violation of lepton flavor, the
observation of lepton flavor violating processes does not require
neutrinos to be massive. In fact, there are many examples of the
latter, the 2HDM being the simplest one. Indeed, in the model
presented in Sec.~\ref{subsec:notation-2HDM}, the most general 2HDM,
neutrino masses are exactly zero but processes such as $\ell_i \to
\ell_j \gamma$ or $h \to \ell_i \ell_j$ are perfectly
possible. Similarly, neutrino masses vanish in the Minimal
Supersymmetric Standard Model, but LFV processes are indeed induced if
the slepton soft mass contain off-diagonal entries. Other examples are
also known, see for instance \cite{Bernabeu:1987gr}.

There are, however, many neutrino mass models that require the
introduction of a second Higgs doublet, and these may potentially lead
to an interesting connection between the generation of neutrino masses
and HLFV. One of the most popular examples of this link is the Zee
model~\cite{Zee:1980ai}, a setup that induces neutrino masses at the
1-loop level.~\footnote{See \cite{Cai:2017jrq} for a comprehensive
  review of radiative neutrino mass models and their phenomenology.}
This model can actually be regarded as an extension of the general
2HDM with the addition of a singly-charged scalar,
\begin{equation}
k \sim \left( {\bf 1}, {\bf 1}, 1 \right) \, .
\end{equation}
If both lepton doublets couple to leptons, as in the type-III 2HDM,
the simultaneous presence of the Yukawa term $f \, \bar \ell^c \, \ell
\, k$ and the trilinear scalar potential term $\mu \, \varphi_1 \,
\varphi_2 \, k^\dagger$ breaks lepton number in two units, thus
inducing Majorana neutrino masses. Therefore, the model contains all
the ingredients to induce neutrino masses and observable HLFV
rates. Interestingly, these two consequences of the Zee model are
connected in a non-trivial way. Assuming for simplicity $f^{e\mu} =
0$, neglecting the electron mass compared to the muon and tau masses,
and keeping the term proportional to the muon mass in the (3,3)
element to get three massive neutrinos, the neutrino mass matrix is
given by~\cite{Herrero-Garcia:2017xdu}
\begin{equation*}
\mathcal M_\nu = A \, s_{2 \delta} \, m_\tau \, \left(\begin{array}{ccc}
-2 f^{e\tau} \rho_e^{\tau e} & -f^{e\tau} \rho_e^{\tau\mu} - f^{\mu\tau} \rho_e^{\tau e} & \frac{\sqrt{2} s_{\beta_\tau}\,m_\tau}{v} f^{e\tau} - f^{e\tau} \rho_e^{\tau\tau}\\
-f^{e\tau} \rho_e^{\tau\mu} - f^{\mu\tau} \rho_e^{\tau e} & -2 f^{\mu\tau} \rho_e^{\tau\mu} & \frac{\sqrt{2} s_{\beta_\tau} m_\tau}{v} f^{\mu\tau} - f^{\mu\tau} \rho_e^{\tau\tau}\\
\frac{\sqrt{2} s_{\beta_\tau}\,m_\tau}{v} f^{e\tau} - f^{e\tau} \rho_e^{\tau\tau} & \frac{\sqrt{2} s_{\beta_\tau} m_\tau}{v} f^{\mu\tau} - f^{\mu\tau} \rho_e^{\tau\tau} & 2 \frac{m_\mu}{m_\tau}f^{\mu\tau} \rho_e^{\mu\tau} \end{array} \right) \,,
\end{equation*}
where $A$ is a dimensionless combination of model parameters,
containing the corresponding loop function, $s_{2 \delta} \propto \mu$
quantifies the mixing in the charged scalar sector and we denote
$s_{\beta_\tau} = \sin \beta_\tau$. The $\rho_e$ matrix was introduced
in Eq. \eqref{eq:Y2HDM}. We see that in order to accommodate the
measured leptonic mixing angles (see for instance the global fit
\cite{deSalas:2017kay}) both $\rho_e^{\tau \mu}$ and $\rho_e^{\tau e}$
must be different from zero. Therefore, the Zee model leads to
correlations between the leptonic mixing angles and the $h \to \tau
\mu$ and $h \to \tau e$ rates. These can be used to set a
\textit{lower} limit on the HLFV
rates~\cite{Herrero-Garcia:2017xdu}. For instance, one finds
\begin{equation}
\BR(h \to \tau \mu) \gtrsim 10^{-6} \, \left( 10^{-7} \right)
\end{equation}
for normal (inverted) neutrino mass ordering. We refer to
\cite{Herrero-Garcia:2017xdu}, where a detailed exploration of the
parameter space of the Zee model is performed, concluding that the
model can accommodate large HLFV rates, even exceeding the current
bounds. Similar findings were recently found in \cite{Nomura:2019dhw},
where a Zee model supplemented with a flavor-dependent $\rm U(1)$
symmetry was considered.

There are other neutrino mass models including a second Higgs
doublet. For instance, in left-right symmetric models
\cite{Pati:1974yy,Mohapatra:1974gc,Senjanovic:1975rk,Mohapatra:1979ia,Mohapatra:1980yp}
one usually introduces a scalar field that is a doublet of both $\rm
SU(2)_L$ and $\rm SU(2)_R$. This bidoublet can be denoted by $\Sigma$
and decomposed as
\begin{equation}
\Sigma = \left( \begin{array}{cc}
\varphi_1^0 & \varphi_2^+ \\
\varphi_1^- & \varphi_2^0 \end{array} \right) \, .
\end{equation}
The scalar representation $\Sigma$ has two gauge invariant Yukawa
couplings to leptons and can be regarded at energies below the $\rm
SU(2)_R$ breaking scale as a pair of $\rm SU(2)_L$ doublets. However,
these setups cannot be identified with a type-III 2HDM since the
left-right symmetry require the two lepton Yukawa matrices to be
Hermitian, thus strongly restricting the allowed parameter
space. Furthermore, current limits on quark flavor violation require
the second CP-even mass eigenstate, $H$, to have a large mass, $m_H
\gtrsim 25$ TeV~\cite{Zhang:2007da}, suppressing all HLFV
effects. Therefore, the minimal left-right models would have to be
enlarged with additional scalar fields in order to be able to provide
large HLFV rates~\cite{Herrero-Garcia:2016uab}.

\section{Summary and discussion}
\label{sec:conclusions}

In this mini-review we have discussed Higgs lepton flavor violating
decays, such as $h \to \ell_i \ell_j$, in the context of the general
2HDM. After motivating this scenario with some model-independent
considerations, we have explicitly shown that the 2HDM can indeed
allow for large HLFV rates while being in perfect agreement with the
experimental constraints at low and high energies. A possible
connection to the mechanism behind the generation of neutrino masses
is also discussed.

The 2HDM must face many stringent constraints in order to generate
large HLFV rates. The first tension comes from existing measurements
of the Higgs boson couplings to fermions and gauge bosons, which
already place bounds on the parameter that controls the mixing between
the two CP-even scalars in the model, $\sin (\beta - \alpha)$. This
must necessarily deviate from $1$ in order to allow for non-standard
Higgs decays, but not too much in order to be in agreement with the
constraints on the Higgs boson couplings. The dipole transitions
$\ell_i \to \ell_j \gamma$ also set very strong limits on the LFV
parameters, and these cannot be avoided by any flavor
symmetry. However, the associated operators turn out to be suppressed
in the 2HDM compared to the operators leading to HLFV, thus increasing
the chances to have observable effects of the latter. Taking all these
constraints into account, as well as those from electroweak precision
data or direct LHC searches, we find that the 2HDM can accommodate
HLFV rates arbitrarily close to the current limits, hence making them
a very attractive way to search for NP.

The determination of the properties of the recently discovered Higgs
boson is among the current priorities for the particle physics
community. The increasingly precise measurements of the Higgs
couplings and decay rates might eventually reveal a deviation from the
SM expectations and hint towards the existence of NP. In particular,
the search for HLFV has just started. A positive signal could shed
light on questions as central as the flavor puzzle or the fundamental
nature of electroweak symmetry breaking.

\section*{Acknowledgements}

I am very grateful to my collaborators in the subjects discussed in
this review. In particular, I thank Diego Aristiz\'abal Sierra for
many long and enjoyable discussions on Higgs lepton flavor violating
decays. I also thank Luca Fiorini for drawing my attention to updated
results on $h \to \ell_i \ell_j$ by the ATLAS collaboration. I
acknowledge financial support from the Spanish grants SEV-2014-0398
and FPA2017-85216-P (AEI/FEDER, UE) and SEJI/2018/033 (Generalitat
Valenciana) and the Spanish Red Consolider MultiDark
FPA2017‐90566‐REDC.

\appendix

\section{Contributions to $\tau \to \mu \gamma$ in the type-III 2HDM}
\label{app:contri}

The most relevant 1- and 2-loop contributions to $\tau \to \mu \gamma$
in the type-III 2HDM were computed in
Ref.~\cite{Davidson:2010xv}. Splitting the form factor $A$ as in
Eq.~\eqref{eq:contri}, these are given by
\begin{align}
A_1 &= \sqrt{2} \, \sum_\phi   
 \frac{ g_{\phi \mu \tau} \, g_{\phi \tau \tau}}{m_{\phi}^2}
\left( \ln \frac{m_{\phi}^2}{m_\tau^2} - \frac{3}{2} \right) \, , \\
A_2^{t,b} &= 6 \sum_{\phi,f} g_{\phi \mu \tau} \, g_{\phi ff} \,
\frac{Q_f^2 \, \alpha}{\pi \, m_\tau \, m_f}
\, f_\phi(r_f) \, , \nonumber \\
A_2^W &= - \sum_{\phi = h,H} g_{\phi \mu \tau} \, C_{\phi WW} \,
 \frac{g \, \alpha}{2 \, \pi \, m_\tau \, m_W}
\left[ 3 f_\phi(r_W)  + \frac{23}{4} 
g (r_W) + \frac{3}{4} 
h (r_W)  + \frac{f_\phi(r_W)
- g(r_W)}{2 \, r_W} \right] \, . \nonumber
\end{align}
Here $\phi = h, H, A$ and $f$ = $t,b$ and we have defined the ratios
\begin{equation}
r_f = \frac{m_f^2}{m_\phi^2} \quad \text{and} \quad r_W = \frac{m_W^2}{m_\phi^2} \, .
\end{equation}
In the derivation of these contributions, the charged lepton masses
have been neglected whenever possible. The expressions for the
$g_{\phi ff^\prime}$ and $C_{\phi W W}$ couplings are given in
Sec. \ref{subsec:notation-2HDM}. Finally, the loop functions
introduced in the previous expressions are given by
\cite{Chang:1993kw}
\begin{align}
f_A(z) \equiv g(z) &= \frac{z}{2} \int_0^1 dx  \frac{1}{ x(1-x)-z } \ln \frac{x(1-x)}{z} \, , \\
f_{h,H}(z) &= \frac{z}{2} \int_0^1 \, dx \, \frac{ (1-2x(1-x))}{x(1-x)-z} \ln\frac{x(1-x)}{z} \, , \\
h(z) &= - \frac{z}{2} \int_0^1 \frac{dx}{x(1-x)-z} \left[1 - \frac{z}{x(1-x)-z} \ln\frac{x(1-x)}{z} \right] \, .
\end{align}

\providecommand{\href}[2]{#2}\begingroup\raggedright\endgroup

\end{document}